\documentclass[12pt]{article}
\usepackage{epsfig}
\usepackage{graphicx}
\usepackage{amsmath}

\begin{document}

 \newcommand{\beq}{\begin{equation}}
\newcommand{\eeq}{\end{equation}}
\newcommand{\bea}{\begin{eqnarray}}
\newcommand{\eea}{\end{eqnarray}}
\newcommand{\beqn}{\begin{eqnarray}}
\newcommand{\eeqn}{\end{eqnarray}}
\newcommand{\beas}{\begin{eqnarray*}}
\newcommand{\eeas}{\end{eqnarray*}}
\newcommand{\defi}{\stackrel{\rm def}{=}}
\newcommand{\non}{\nonumber}
\newcommand{\bquo}{\begin{quote}}
\newcommand{\enqu}{\end{quote}}
\newcommand{\qt}{\tilde q}
\newcommand{\m}{\tilde m}
\newcommand{\trho}{\tilde{\rho}}
\newcommand{\tn}{\tilde{n}}
\newcommand{\tN}{\tilde N}
\newcommand{\gsim}{\lower.7ex\hbox{$\;\stackrel{\textstyle>}{\sim}\;$}}
\newcommand{\lsim}{\lower.7ex\hbox{$\;\stackrel{\textstyle<}{\sim}\;$}}


\def\de{\partial}
\def\Tr{ \hbox{\rm Tr}}
\def\const{\hbox {\rm const.}}
\def\o{\over}
\def\im{\hbox{\rm Im}}
\def\re{\hbox{\rm Re}}
\def\bra{\langle}\def\ket{\rangle}
\def\Arg{\hbox {\rm Arg}}
\def\Re{\hbox {\rm Re}}
\def\Im{\hbox {\rm Im}}
\def\diag{\hbox{\rm diag}}


\def\QATOPD#1#2#3#4{{#3 \atopwithdelims#1#2 #4}}
\def\stackunder#1#2{\mathrel{\mathop{#2}\limits_{#1}}}
\def\stackreb#1#2{\mathrel{\mathop{#2}\limits_{#1}}}
\def\Tr{{\rm Tr}}
\def\res{{\rm res}}
\def\Bf#1{\mbox{\boldmath $#1$}}
\def\balpha{{\Bf\alpha}}
\def\bbeta{{\Bf\beta}}
\def\bgamma{{\Bf\gamma}}
\def\bnu{{\Bf\nu}}
\def\bmu{{\Bf\mu}}
\def\bphi{{\Bf\phi}}
\def\bPhi{{\Bf\Phi}}
\def\bomega{{\Bf\omega}}
\def\blambda{{\Bf\lambda}}
\def\brho{{\Bf\rho}}
\def\bsigma{{\bfit\sigma}}
\def\bxi{{\Bf\xi}}
\def\bbeta{{\Bf\eta}}
\def\d{\partial}
\def\der#1#2{\frac{\d{#1}}{\d{#2}}}
\def\Im{{\rm Im}}
\def\Re{{\rm Re}}
\def\rank{{\rm rank}}
\def\diag{{\rm diag}}
\def\2{{1\over 2}}
\def\ntwo{${\mathcal N}=2\;$}
\def\nfour{${\mathcal N}=4\;$}
\def\none{${\mathcal N}=1\;$}
\def\ntwot{${\mathcal N}=(2,2)\;$}
\def\ntwoo{${\mathcal N}=(0,2)\;$}
\def\x{\stackrel{\otimes}{,}}

\def\ba{\beq\new\begin{array}{c}}
\def\ea{\end{array}\eeq}
\def\be{\ba}
\def\ee{\ea}
\def\stackreb#1#2{\mathrel{\mathop{#2}\limits_{#1}}}

\def\Tr{{\rm Tr}}
\newcommand{\cpn}{CP$(N-1)\;$}
\newcommand{\wcpn}{wCP$_{N,\tilde{N}}(N_f-1)\;$}
\newcommand{\wcpd}{wCP$_{\tilde{N},N}(N_f-1)\;$}
\newcommand{\vp}{\varphi}
\newcommand{\pt}{\partial}
\newcommand{\ve}{\varepsilon}
\renewcommand{\theequation}{\thesection.\arabic{equation}}

\setcounter{footnote}0

\vfill

\begin{titlepage}

\begin{flushright}
FTPI-MINN-13/07, UMN-TH-3139/13\\
\end{flushright}

\vspace{1mm}

\begin{center}
{  \Large \bf  
\boldmath{Hybrid $r$-Vacua  
 in  
 \boldmath{\ntwo} 
Supersymmetric \\[1mm]
QCD: Universal Condensate Formula
}}

\vspace{10mm}

 {\large \bf    M.~Shifman$^{\,a}$ and \bf A.~Yung$^{\,\,a,b}$}
\end {center}

\begin{center}

$^a${\it  William I. Fine Theoretical Physics Institute,
University of Minnesota,
Minneapolis, MN 55455, USA}\\
$^{b}${\it Petersburg Nuclear Physics Institute, Gatchina, St. Petersburg
188300, Russia
}
\end{center}

 \vspace{10mm}

\begin{center}
{\large\bf Abstract}
\end{center}

We derive an exact unified formula for all condensates (quark and mono\-pole)
in the hybrid $r$ vacua in 
\ntwo  supersymmetric QCD slightly deformed by a $\mu{\cal A}^2$ term.
The gauge group is assumed to be  U$(N)$   and the number of the quark flavors $N_f$ subject to the
condition
$N< N_f < 2N$. In the $r$ vacua 
  $r$ quarks and $N-r-1$ monopoles from non-overlapping subgroups of U$(N)$ 
  develop vacuum expectation values (VEVs)
   ($r<N$).
We  then briefly review possible dynamical regimes  (confinement, screening, and ``instead of confinement")
 in the  hybrid $r$ vacua in $\mu$-deformed \ntwo SQCD (the small-$\mu$ limit).

\vspace{2cm}

\end{titlepage}

 \newpage



\section {Introduction }
\label{intro}
\setcounter{equation}{0}

The main goal of this paper is to derive a unified formula for the quark and monopole vacuum condensates 
in an arbitrary $r$ vacuum in \ntwo supersym\-metric QCD (SQCD) in terms of the roots of the Seiberg-Witten curve
\cite{SW1}. Following Seiberg and Witten we deform  \ntwo SQCD by a small mass term $\mu$
for the adjoint field. 
We will show that all the condensates reduce to effective parameters $\xi_P$,
\beq
\xi_P=-2\sqrt{2}\,\mu\,\sqrt{(e_P-e_N^{+})(e_P-e_N^{-})}
\label{xiv}
\eeq
where the subscript $P=1,..., N-1$ marks the appropriate condensates (quark or monopole), 
$e_1,\,e_2,\, ...,  e_{N-1}$ are the double roots of the  Seiberg-Witten curve corresponding to 
the quark and monopole condensation, while $e_N^{\pm}$ are two unpaired roots present in any $r<N$ vacuum in the case of the $\mu\, \Tr\, {\cal A}^2$ perturbation. 
If $P$ lies in the interval $[1, \, r]$, Eq. (\ref{xiv}) describes the  quark 
vacuum expectation values (VEVs) \cite{SYrvacua}, while
for $r+1\leq P\leq N-1$ it gives the monopole VEVs.

For generic values of the quark masses
the theories we discuss support BPS-saturated
non-Abelian magnetic strings
 \cite{HT1,ABEKY,SYmon,HT2}. These strings confine monopoles. The tensions of these strings are  \cite{SYrev,SYfstr}
 \beq
T_P=2\pi|\xi_P|\, , \qquad P=1, ..., r\,.
\label{mten}
\eeq
For $r+1\leq P\leq N-1$ the same expression gives the tensions of the Abelian electric strings, which confine quarks.
The value of the $P$-th condensate is $\xi_P/2$ (see below for a more precise definition). 

Let us briefly outline our basic model (a more detailed description and all relevant notation can be found in our previous
original publications \cite{SYmon,SYfstr} and the
review papers \cite{SYrev}).

The gauge group of  \ntwo SQCD under consideration is  U$(N)$. We introduce $N_f$ quark flavors  ($N< N_f < 2N$) 
endowed with mass terms and then perturb \ntwo SQCD by 
a small mass term $\mu{\cal A}^2$ for the adjoint matter (part of the \ntwo gauge supermultiplet). 

At generic quark masses this theory has a number of isolated vacua where
$r$ flavors of (s)quarks condense, $r\le N$ (the so called $r$ vacua). The $r=N$ vacuum, with the maximal possible   number of condensed quarks, was 
studied more than others (for a  review see \cite{SYrev}). Non-Abelian flux tubes (strings)    confining monopoles were shown to exist \cite{HT1,ABEKY,SYmon,HT2} in this vacuum, see 
\cite{Trev,Jrev,SYrev} for extensive reviews. Massless $r$-vacua with $r<N$ were studied in 
\cite{APS,CKM} in the SU$(N)$ version of the theory.\footnote{If quark mass terms vanish certain $r$
vacua  
coalesce, and the Higgs branches develop from the common roots. The $r<N$ vacua  
 correspond to roots of the nonbaryonic Higgs branches, while the $r=N$ vacuum  to 
a root of the baryonic Higgs branch in the SU$(N)$ theory \cite{APS}.
We consider nonvanishing, nondegenerative quark masses.}

Extensions to  U$(N)$ were discussed recently for the 
$r>N_f/2$ and,  in particular,  $r=N-1$  and $r=N$ cases
 \cite{SYdual,SYN1dual,SYrvacua}. 
Confinement of monopoles at weak coupling was demonstrated to survive in the  strong coupling regime
at small values of the quark VEVs given by  $\xi/2\sim \mu m$ where $m$ is a typical quark mass. The latter was described in terms of the so-called $r$-duality and was found  to be  an ``instead-of-confinement'' phase: the screened quarks decay into monopole-antimonopole pairs with the monopoles 
confined by non-Abelian strings.
One of the results of \cite{SYrvacua} was the expression for the 
quark condensates in the low-energy theory in terms of the roots of the Seiberg-Witten curve, see Eq. (\ref{xiv}).
In this paper we continue this line of research and consider the monopole $r=0$ as well as hybrid $r$
 vacua  with $r$ quarks and $(N-r-1)$ monopoles (from the orthogonal subgroups of U$(N)$) condensing.\footnote{
A certain aspect of the large-$\mu$ limit was not quite adequately treated in \cite{SYrvacua}. This will be corrected in a separate publication. In the present paper we limit ourselves to the small-$\mu$ limit.}
Equation (\ref{xiv}) proves to be valid for all condensates in all vacua. Although our derivation will be carried out in particular examples the assertion is universal.
 
 The paper is organized as follows. In Sec. \ref{bulk} we discuss the $r$-vacuum 
 structure and review Eq. (\ref{xiv}) for $r>N_f/2$.  In Sec. \ref{monopolevac} we present a detailed analysis of the  monopole ($r=0$) vacuum
and derive Eq. (\ref{xiv}) in this case. As a byproduct we observe that  Eq. (\ref{xiv}) reproduces
the famous sine formula for the  string tensions \cite{DS} in the limit of large quark masses,
when the theory under consideration  reduces to pure gauge theory.\footnote{For a related discussion see
\cite{Armoni}.}
Section  \ref{hybridvac} is devoted to
the hybrid $r$-vacua with $r<N_f/2$.  Equation (\ref{xiv}) for the quark and monopole condensates is derived in certain examples.
Finally, 
Section \ref{conf} presents an overall  picture of confinement and screening in the  hybrid $r$ vacua. In Sec. \ref{five} we also
summarize various phases exhibiting themselves  in different $r$ vacua. Appendix contains details pertinent to the VEVs calculation  in a hybrid vacuum.

\section { \boldmath{$\mu$}-Deformed  SQCD:  vacuum structure}
\label{bulk}
\setcounter{equation}{0}

\subsection{The model}

 In the absence
of  deformation the model under consideration is \ntwo  SQCD
 with $N_f$ massive quark hypermultiplets. 
 We assume that
$N_f>N$ but $N_f< 2 N$ where $N$ refers to the gauge group, U($N$). 
The latter inequality ensures our theory to be asymptotically free. 
In addition, we will introduce a small mass term $\mu{\cal A}^2$ 
for the adjoint matter breaking \ntwo supersymmetry down to \none. 

The field content is as follows. In addition to the SU($N)$ and U(1)
 \ntwo gauge supermultiplets we have 
$N_f$ quark multiplets  consisting
of   the complex scalar fields
$q^{kA}$ and $\tilde{q}_{Ak}$ (squarks) and
their  fermion superpartners --- all in the fundamental representation of 
the SU$(N)$ gauge group.
Here $k=1,..., N$ is the color index
while $A$ is the flavor index, $A=1,..., N_f$. We will treat $q^{kA}$ and $\tilde{q}_{Ak}$
as rectangular matrices with $N$ rows and $N_f$ columns. 

 The  superpotential of the undeformed theory is 
 \beq
{\mathcal W}_{{\mathcal N}=2} = \sqrt{2}\,\sum_{A=1}^{N_f}
\left( \frac{1}{ 2}\,\tilde q_A {\mathcal A}
q^A +  \tilde q_A {\mathcal A}^a\,T^a  q^A + m_A\,\tilde q_A q^A\right)\,,
\label{superpot}
\eeq
where ${\mathcal A}$ and ${\mathcal A}^a$ are  chiral \none superfields, the ${\mathcal N}=2$
superpartners of the gauge bosons, while $m_A$ are the quark mass terms.
Then we add a single trace deformation  
\beq
{\mathcal W}_{{\rm br}}=
  \mu\,{\rm Tr}\,\Phi^2,
\label{msuperpotbr}
\eeq
where
\beq
\Phi=\frac12\, {\mathcal A} + T^a\, {\mathcal A}^a\,,
\label{Phi}
\eeq
and $T^a$ stand for the SU$(N)$ generators. 
Generally speaking, (\ref{msuperpotbr}) breaks\footnote{For small $\mu$ and    equal quark masses
(\ref{msuperpotbr}) reduces to the Fayet-Iliopoulos $F$-term \cite{FI} which does not break \ntwo supersymmetry, see \cite{HSZ,VY,SYfstr}.} \ntwo
supersymmetry down to {\none}.
We assume the deformation (\ref{msuperpotbr}) to be weak,
\beq
|\mu | \ll \Lambda\, ,
\label{smallmu}
\eeq
where $\Lambda$ is the scale of the \ntwo  theory. Thus, we consider
the theory close to its \ntwo limit.

\subsection{Vacua}
The number of isolated $r=N$ vacua is  
\beq
{\cal N}_{r=N} = C_{N_f}^{N}= \frac{N_f!}{N!(N_f-N)!}\,.
\label{numNvac}
\eeq
This is the maximal number  of quark fields that can develop
VEVs, see \cite{SYrev}. All gauge   bosons are completely Higgsed and the
theory is in the color-flavor locking phase (assuming quark masses to be close to each other). 
The quark VEVs are determined by
 $\xi_P$'s ($P=1,..., N$)  of the order of
  $ \mu m_P$. For large values of $\xi$ the theory is at weak coupling
and can be studied semiclassically. In particular, non-Abelian strings  are known to exist
which confine monopoles \cite{HT1,ABEKY,SYmon,HT2}.

If we reduce  $\xi$   the theory undergoes 
 a crossover transition from weak to strong coupling regime, 
described in terms of a weakly coupled infrared-free  dual
 theory  \cite{SYdual} with the  U$(\tN)$ gauge group and  $N_f$ light quark-like dyon flavors, $\tN = N_f-N$.
 The dyon condensation leads to confinement 
of monopoles too. 
The quarks and gauge bosons of the original theory are in  the
``instead-of-confinement'' phase \cite{SYdual,SYrvacua}.

The number of the $r$ vacua\footnote{Our definition of $r$ refers to the large quark mass
 domain. In fact,
 effectively $r$ depends on the quark masses, see \cite{SYvsS}.} with $r<N$  is \cite{CKM}
\beq
{\cal N}_{r<N}=\sum_{r=0}^{N-1} \,(N-r)\,C_{N_f}^{r}= \sum_{r=0}^{N-1}\, (N-r)\,\frac{N_f!}{r!(N_f-r)!}\,,
\label{nurvac}
\eeq
representing the number of choices one can pick up $r$ condensing quarks out of $N_f$ quarks times the
Witten index  in the classically unbroken SU$(N-r)$ pure gauge theory. 

Consider a
 particular vacuum in which the first $r$ quarks develop VEVs. 
We denote it as ($1, ... ,\, r$). Quasiclassically, at large masses,  the adjoint scalar VEVs   are  
\beq
\left\langle \Phi\right\rangle \approx - \frac1{\sqrt{2}}\,
{\rm diag}\left[m_1,...,m_r,0,...,0 
\right],
\label{avevr}
\eeq
where  the last $(N-r)$ entries   classically vanish.
In quantum theory  the vanishing entries become of the order of $\Lambda$, generally speaking.
The classically unbroken U$(N-r)$ gauge sector gets Higgsed  through the Seiberg--Witten mechanism \cite{SW1},
first down to U(1)$^{N-r}$ and then almost completely by condensation of $(N-r-1)$ monopoles. A single
 U(1) factor remains unbroken,
as the monopoles are charged with respect to 
the Cartan generators of the SU$(N-r)$ group.

The presence of the unbroken U(1)$^{\rm unbr}$ symmetry
  makes the $r<N$ vacua qualitatively different from the $r=N$ vacuum:
  the latter has 
no massless gauge bosons. According to \cite{Cachazo2},  these sets of vacua belong to 
two different ``branches.''

The low-energy theory in the $r$ vacuum  has the gauge group
\beq
{\rm U}(r)\times {\rm U}(1)^{N-r}\,,
\label{legaugegroup}
\eeq
 with $N_f$ quark flavors  charged
under the U$(r)$ factor and $(N-r-1)$ monopoles charged under the U(1) factors. 

\subsection{\boldmath{$r> N_f/2$}}
\label{larger}

For $r> N_f/2$ and large $\xi$ the SU$(r)$ non-Abelian
quark sector is at weak coupling since it is asymptotically free.\footnote{The opposite case $r< N_f/2$
is discussed in Sec. \ref{hybridvac}.}
The action of this  theory is presented in \cite{SYrvacua} for a particular example, the $r=N-1$ vacuum.
The  quark condensates can be read-off from the superpotentials (\ref{superpot}) and (\ref{msuperpotbr}) using 
(\ref{avevr}). They are 
\beqn
\langle q^{kA}\rangle &=& \langle\bar{\tilde{q}}^{kA}\rangle=\frac1{\sqrt{2}}\,
\left(
\begin{array}{cccccc}
\sqrt{\xi_1} & \ldots & 0 & 0 & \ldots & 0\\
\ldots & \ldots & \ldots  & \ldots & \ldots & \ldots\\
0 & \ldots & \sqrt{\xi_r} & 0 & \ldots & 0\\
\end{array}
\right),
\nonumber\\[4mm]
k&=&1,..., r\,,\qquad A=1,...,N_f\, .
\label{qvevr}
\eeqn
The first  $r$ parameters $\xi$ in the  quasiclassical  approximation are 
\beq
\xi_P \approx 2\;\mu m_P,
\qquad P=1,..., r.
\label{xiclass}
\eeq
In quantum theory the parameters $\xi_P$   determining  the  quark condensates are connected with the roots of the Seiberg-Witten curve \cite{SYfstr,SYN1dual,SYrvacua} which in the theory at hand
 takes the form \cite{APS}
\beq
y^2= \prod_{P=1}^{N} (x-\phi_P)^2 -
4\left(\frac{\Lambda}{\sqrt{2}}\right)^{2N-N_f}\, \,\,\prod_{A=1}^{N_f} \left(x+\frac{m_A}{\sqrt{2}}\right).
\label{curve}
\eeq
Here $\phi_P$ are gauge invariant parameters on the Coulomb branch. Semiclassically,
\beq
\Phi \approx 
{\rm diag}\left[\phi_1,...,\phi_N\right].
\eeq
In the  $r<N$ vacuum (more exactly, in the ($1, ... ,\, r$) vacuum) we have
\beq
\phi_P \approx -\frac{m_P}{\sqrt{2}},\qquad P=1, ... ,\, r\,, \qquad 
\phi_P \sim \Lambda_{{\mathcal N}=2},\qquad P=r+1, ... ,\, N\,
\label{classphi}
\eeq
in the large $m_A$ limit, see (\ref{avevr}).

To identify the $r<N$ vacuum in terms of the curve (\ref{curve}) it is necessary to find
such values of $\phi_P$ which ensure the Seiberg-Witten curve to have $N-1$ double roots, with
$r$ parameters $\phi_P$  determined by the quark masses in the semiclassical limit, see (\ref{classphi}).
The above $N-1$ double roots will be associated with the $r$ condensed quarks and $(N-r-1)$ condensed monopoles --
 altogether $N-1$ condensed states.
 
In contrast, in the $r=N$ vacuum the maximal possible number of condensed states (quarks) 
in the U$(N)$ theory is $N$. As was mentioned, this difference is related to the 
the unbroken U(1)$^{\rm unbr}$ gauge group
in the $r<N$ vacua \cite{Cachazo2}. In the classically unbroken (after the quark condensation) U$(N-r)$ gauge group,
$N-r-1$ monopoles condense at a quantum level,  breaking the non-Abelian SU$(N-r)$ subgroup. One U(1)
factor remains unbroken because the monopoles are not coupled  to this U(1).

Thus in the $r<N$ vacua with the quadratic deformation superpotential (\ref{msuperpotbr}) the Seiberg--Witten curve factorizes \cite{CaInVa},
\beq
y^2
=\prod_{P=1}^{r} (x-e_P)^2\,\prod_{K=r+1}^{N-1} (x-e_K)^2\,(x-e_N^{+})(x-e_N^{-})\,.
\label{rcurve}
\eeq
The first $r$ double roots in the large mass limit are given  by the
mass parameters, $\sqrt{2}e_P\approx  -m_P$, $P=1, ... , r$. Other $(N-r-1)$ double roots associated with light monopoles are much smaller and determined by $\Lambda$.
The last two  roots  are also much smaller. 

 For the single-trace deformation superpotential (\ref{msuperpotbr}) 
 the sum of the unpaired roots vanishes \cite{CaInVa},
\beq
e_N^{+} + e_N^{-}=0\,.
\label{DijVafa}
\eeq
The root $e_N^{+}$ determines the value of the gaugino condensate \cite{Cachazo2}.

Now, Eq. (\ref{xiv}) was derived in one of our previous papers \cite{SYrvacua}
for the case of the quark condensate namely, for $P=1,...,r$.



In the remainder of this paper we demonstrate that the monopole condensates in the monopole vacuum ($r=0$) or hybrid $r$ vacua  are also determined by the same
formula with the replacement of the quark double roots by the monopole double roots, 
so that the subscript   $P$ in (\ref{xiv}) can  run
over monopole double roots $P=(r+1),..., (N-1)$ too. Thus Eq. (\ref{xiv}) is very general and determines
VEVs of any condensed field independently of its nature.

\section {\boldmath{$r=0$}: the  monopole vacuum}
\label{monopolevac}
\setcounter{equation}{0}

In this section we  consider the  monopole vacuum with $r=0$  and show that the monopole condensates are still given 
by Eq. (\ref{xiv}). Then,  we demonstrate that for the above  monopole vacuum 
(in the limit of large quark masses, i.e. when the theory at hand reduces to pure gauge theory)
 Eq. (\ref{xiv}) gives the famous sine formula for the monopole VEVs and, hence, the electric string tensions \cite{DS}.

\subsection{Monopole VEVs}

Consider the simplest example: the  $r=0$ vacuum in  U(2) SQCD with
 $N_f$ quark  flavors.
It is a straightforward generalization  of the SU(2) theory studied in \cite{SW1,SW2}.
The low-energy gauge group is U(1)$\times$ U(1) where the first U(1) factor is associated with, say, the $\tau_3$
generator of SU(2). In this case the light matter sector consists of one monopole singlet $M$ and $\tilde{M}$ charged with
respect to the first U(1) factor \cite{SW1}.  
The relevant $F$-terms in the scalar potential are
\beqn
V(M,\tilde{M},a^D_3,a) &=&
2g^2_D\left| \tilde{M} M
+\frac{\mu}{\sqrt{2}}\,\,\frac{\pt u_2}{\pt a^D_3}\right|^2+
g^2_1\left| 
\mu\,\frac{\pt u_2}{\pt a}\right|^2
\nonumber\\[3mm]
&+&
2\left|a^D_3 M\right|^2 + 2\left|a^D_3\bar{\tilde{M}}\right|^2 +\cdots \, ,
\label{potr=0}
\eeqn
where we denote the light adjoint scalar of the dual gauge multiplet
associated with $\tau_3$ by $a^D_3$, while $a$ stands for the neutral scalar in the U(1) gauge multiplet of U(2). 
The corresponding coupling 
constants are $g_D$ and $g_1$, respectively. We also define
\beq
u_k= \left\langle {\rm Tr}\left(\frac12\, a + T^a\, a^a\right)^k\right\rangle, \qquad k=1, ..., N\,.
\label{u}
\eeq
Thus, the deformation superpotential (\ref{msuperpotbr}) is proportional to $u_2$.
From the potential (\ref{potr=0}) it is easy to derive for the monopole vacuum 
\beqn
\langle  \tilde{M} M \rangle =
-\frac{\mu}{\sqrt{2}}\,\,\frac{\pt u_2}{\pt a^D_3}\,; \qquad \frac{\pt u_2}{\pt a} = 0, \quad a^D_3=0\,.
\label{r=0vac}
\eeqn
The Seiberg-Witten curve  in this case factorizes as follows:
\beq
y^2=(x-e_1)^2\,(x-e_2^{+})(x-e_2^{-}),
\label{curveN=2}
\eeq
 see (\ref{rcurve}). Here the double root at $x=e_1$ corresponds to a single condensed monopole in 
the $r=0$
 vacuum, while
two other roots (subject to the condition (\ref{DijVafa})) determine the gaugino condensate.

\begin{figure}[h]
\epsfxsize=6cm
\centerline{\epsfbox{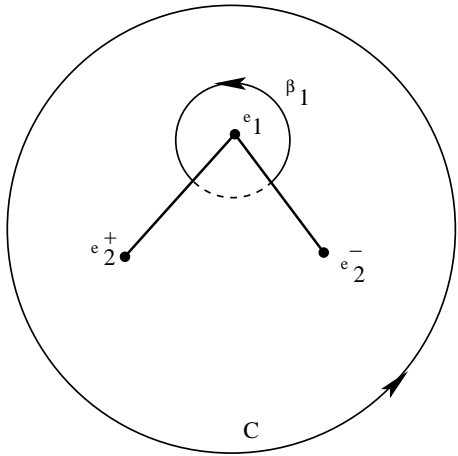}}
\caption{\small $\beta_1$ and $C$-contours in the $x$ plane in the U(2) theory. Solid straight lines denote  cuts. }
\label{figr=0contours}
\end{figure}

The exact solution of the theory on the Coulomb branch relates 
the fields   $a^D_3$ and $a$ to 
contour integrals running along the contours $\beta_1$  in the $x$ plane encircling  the double root $e_1$ and the 
contour $C$ at infinity, see Fig~\ref{figr=0contours}.

Using explicit the expressions from \cite{ArFa,KLTY,ArPlSh,HaOz} and their generalizations to 
the U$(N)$ case \cite{SYfstr}
we arrive at
\beqn
&& \frac{\pt a^D_3}{\pt u_2}= \frac12 \,\frac1{2\pi i}\oint_{\beta_1}  \frac{dx}{y}\,,\,\,
\qquad \frac{\pt a^D_3}{\pt u_1}=  \,\frac1{2\pi i}\oint_{\beta_i} \frac{dx}{y}
\left[x-(e_1+e_2)\right]\,,
\nonumber\\ [3mm]
&&
\frac{\pt a}{\pt u_2}= \frac12 \,\frac1{2\pi i}\oint_{C}  \frac{dx}{y}\,,\,\,
\qquad \frac{\pt a}{\pt u_1}=  \,\frac1{2\pi i}\oint_{C} \frac{dx}{y}
\left[x-(e_1+e_2)\right]\,,
\label{dadu20}
\eeqn
where the variables $u_1$ and $u_2$ are given in Eq. (\ref{u}), 
while 
\beq
e_2=\frac12\left(e_2^{+}+e_2^{-}\right).
\label{eN}
\eeq
In fact, $e_2$ should vanish due to the condition (\ref{DijVafa}). We will see shortly that this is indeed the case.

For the factorized curve (\ref{curveN=2}) the integrals (\ref{dadu20}) can be  easily
evaluated. In particular, the integral along the $\beta_1$ contour is given by its pole contributions. 
This gives 
\beqn
&&
\frac{\pt a^D_3}{\pt u_2} =\frac12\,\frac1{\sqrt{(e_1-e_2^{+})(e_1-e_2^{-})}},\qquad
\frac{\pt a}{\pt u_2} = 0,
\nonumber\\[3mm]
&&
\frac{\pt a^D_3}{\pt u_1} = -\frac{e_2}{\sqrt{(e_1-e_2^{+})(e_1-e_2^{-})}}, \qquad
\frac{\pt a}{\pt u_1} =1\,.
\label{dPhidu}
\eeqn
Inverting this matrix we get
\beq
\frac{\pt u_2}{\pt a^D_3}= 2\sqrt{(e_1-e_2^{+})(e_1-e_2^{-})}, \qquad \frac{\pt u_2}{\pt a}  = 2e_2.
\eeq
Now from (\ref{r=0vac}) we see that indeed
\beq
e_2=0\,,
\label{EN=eN}
\eeq
i.e.  the condition (\ref{DijVafa}) is automatically met. The monopole VEV is\,\footnote{Here we also use the $D$-term condition  requiring $|M|=|\tilde{M}|$.}
 \beq
  \langle M\rangle = \langle\bar{\tilde{M}}\rangle = \sqrt{\frac{\xi_1}{2}} 
 \label{mvevN=2}
 \eeq
 with 
 \beq
 \xi_1= - 2\sqrt{2}\,\mu\, \sqrt{(e_1-e_2^{+})(e_1-e_2^{-})}.
 \eeq
 
We see that the monopole condensate in the $r=0$ vacuum is determined by the same Eq. (\ref{xiv}) in much the same way as 
 the quark condensates,
 see (\ref{qvevr}). Straightforward generalization of this result to arbitrary $N$ gives for for elementary 
 monopole condensates   
\beq
  \langle M_{P(P+1)}\rangle = \langle\bar{\tilde{M}}_{P(P+1)}\rangle = \sqrt{\frac{\xi_P}{2}},  
 \label{mvev}
 \eeq
where the parameters $\xi_P$ are again determined by the general formula (\ref{xiv}) ($P=1,..., (N-1)$). Here $M_{PP'}$ denotes the monopole with the charge given by  the root $\alpha_{PP'}=w_P-w_{P'}$ of the SU$(N)$ algebra with weights $w_P$, $P<P'$.

\subsection{The sine formula}

The famous sine formula for the  $k$-string tensions (and, hence, condensates) was derived in \cite{DS} in the \ntwo limit of pure gluodynamics. The latter can be obtained from our model by tending the quark masses to infinity, where they decouple.

Consider the $r=0$ monopole vacuum in the U$(N)$ gauge theory with heavy quarks, $m_A\to \infty $. 
The Seiberg-Witten 
curve in this case takes the form
\beq
y^2= \prod_{P=1}^{N} (x-\phi_P)^2 -
4\left(\frac{\Lambda_0}{\sqrt{2}}\right)^{2N}\,,
\label{YMcurve}
\eeq
where the scale $\Lambda_0$ is 
\beq
\Lambda_0^{2N}= \Lambda^{2N-N_f}\,\prod_{A=1}^{N_f} m_A\, .
\label{YMLambda}
\eeq
 The corresponding expressions for $\phi_P$'s, double monopole 
roots $e_P$ and two unpaired roots $e_N^{\pm}$ are \cite{DS}
\beqn
\phi_P & = & 2 \cos{\frac{\pi (P-\frac12)}{N}}\,\,\frac{\Lambda_0}{\sqrt{2}},  \qquad P=1,...,N,
\nonumber\\[3mm]
e_P & = & 2 \cos{\frac{\pi P}{N}}\,\,\frac{\Lambda_0}{\sqrt{2}}, \qquad P=1,...,(N-1),
\nonumber\\[3mm]
e_N^{\pm} & = & \pm \, 2\,\,\frac{\Lambda_0}{\sqrt{2}}.
\label{YMe}
\eeqn
Substituting these roots in the formula (\ref{xiv}) we arrive at the following monopole VEVs:
\beq
  \langle \tilde{M}_{P(P+1)} M_{P(P+1)}\rangle = \frac{\xi_P}{2}= -2i\mu \Lambda_0\,\,\sin{\frac{\pi P}{N}}\,,  
 \label{YMmvev}
 \eeq
The same monopole VEVs determine the tensions of the Abelian electric strings, 
\beq
T_P=2\pi|\xi_P|\,,\qquad P=1, ... , N-1\,.
\label{eten}
\eeq
Our general expression (\ref{xiv})
 reproduces the sign formula! The string described by (\ref{YMmvev}) can be viewed   \cite{HSZ} as the so-called ``$k$ strings," see \cite{Armoni} and references therein.

In much the same way as the magnetic non-Abelian strings appearing upon the quark condensation
in the $r$ vacua,
these strings are BPS to the leading order in $\mu$ \cite{HSZ,VY}. These Abelian electric strings
confine quarks.

\section {Hybrid \boldmath{$r$} vacua}
\label{hybridvac}
\setcounter{equation}{0}

As was already mentioned, the low-energy gauge group in the hybrid $r$ vacuum is  (\ref{legaugegroup}),
while the light matter sector consist of $N_f$ quark flavors   charged under the U$(r)$ gauge subgroup, plus $(N-r-1)$
singlet Abelian monopoles. The quarks and monopoles are charged with respect to orthogonal subgroups of U$(N)$.
Hence, they are mutually local (i.e. can be described by a local Lagrangian). 
If in Sec. \ref{larger} we discussed the case $r> N_f/2$,
now we turn to the opposite case $r< N_f/2$.

In these vacua the low-energy theory
is infrared free and it is at  weak coupling once the quark and monopole VEVs  are small. To ensure this
condition we assume   all parameters $\xi_P$ given by (\ref{xiv}) to be  small enough.

For example,
for large and  (almost) equal quark masses the effective scale of the
non-Abelian SU$(r)$ subgroup of (\ref{legaugegroup}) is  
\beq
\Lambda_{SU(r)}^{N_f-2r} =\frac{m^{2(N-r)}}{\Lambda^{2N-N_f}}
\eeq
where $m$ is the common mass, and 
 $|\xi_P| \ll \Lambda_{{\rm SU}(r)}^2$, $P=1,...,r$.
For simplicity here and in Sec. \ref{conf} we assume $m$ to be large and hence
 quarks have only electric color charges. For a discussion of the small mass limit
see Sec. \ref{sum}.
 
\begin{figure}[h]
\epsfxsize=7.5cm
\centerline{\epsfbox{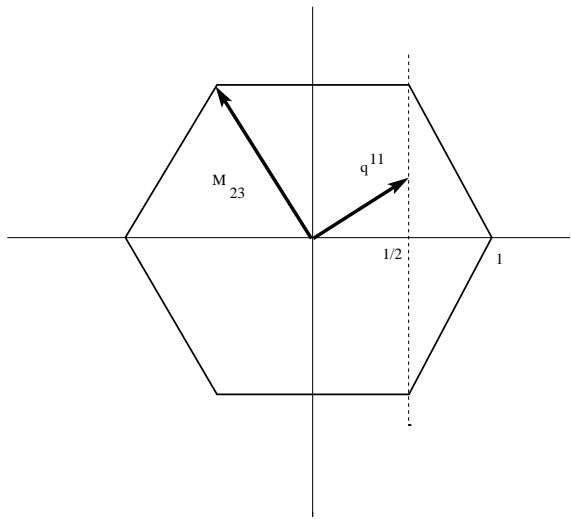}}
\caption{\small Projection of charges (\ref{charges}) of the condensed quark and monopole states onto SU(3) subalgebra of U(3). }
\label{fig:q1m23}
\end{figure}

 As an example we choose for our analysis the $r=1$ vacuum in 
 the U(3) gauge theory with $N_f$ quark flavors.
 The light matter sector consists of a single color component of  $N_f$ 
quark flavors   and a monopole singlet. We can choose
color charges of quarks and monopole as follows (see (\ref{avevr})):
\beqn
\vec{n}_{q^{1A}}=
\left(\frac12,0;\,\frac12,0;\,\frac1{2\sqrt{3}},0\right), \quad \vec{n}_{M_{23}}=
\left(0,0;\,0,-\frac12;\,0,\frac{\sqrt{3}}{2}\right), 
%
\label{charges}
\eeqn
respectively, where we use the notation
\beq
\vec{n}=\left(n_e,n_m;\,n_e^3,n_m^3;\,n_e^8,n_m^8\right),
\label{chargenotation}
\eeq
and
$n_e$ and $n_m$ denote the electric and 
magnetic charges of a given state with respect to the U(1) gauge group. Moreover,  $n_e^3$,
$n_m^3$ and  $n_e^8$, $n_e^8$  stand for the electric and 
magnetic charges  with respect to the Cartan
generators of the SU(3) gauge group.
The charges chosen in (\ref{charges}) correspond to taking the quark charge equal to the weight  $w_1$ and 
monopole charge equal to the orthogonal root $\alpha_{23}=w_2 -w_3$ of SU(3) subgroup of U(3), see Fig.~\ref{fig:q1m23}.

From Eq. (\ref{charges}) we see that the quarks interact with U(1) gauge field 
\beq
A^q_{\mu} = \sqrt{\frac37}\,\left(A_{\mu} + A_{\mu}^3 + \frac1{\sqrt{3}}\, A_{\mu}^8 \right)
\label{Aq}
\eeq
with the charge 
\beq
n_q \equiv \frac12\,\sqrt{\frac73}\,.
\label{nq}
\eeq
At the same time,  the  monopoles interact with the U(1) gauge field
\beq
A^D_{\mu} = \frac12\,\left( A_{\mu}^{D3} + \sqrt{3}\, A_{\mu}^{D8} \right)
\label{AM}
\eeq
with the charge $n_M=1$, while the orthogonal combination
\beq
A^{{\rm unbr}}_{\mu} = \frac{3}{2\sqrt{7}}\,\left(-\frac43 A_{\mu} + A_{\mu}^3 + \frac1{\sqrt{3}}\, A_{\mu}^8 \right)
\label{Aunbr}
\eeq
is the gauge field of the unbroken U(1)$^{{\rm unbr}}$ always present in all $r<N$ vacua. Here $A_{\mu}^{Da}$ denote
dual gauge potentials associated with the Cartan generators of SU(3).

Relevant $F$-terms in the scalar potential of the low-energy theory are
\beqn
V &=&
2g^2_q\left| n_q\,\tilde{q}_{A1} q^{1A}
+\frac{\mu}{\sqrt{2}}\,\,\frac{\pt u_2}{\pt a_q}\right|^2
\nonumber\\[3mm]
&+&
2g^2_M\left| \tilde{M}_{23} M_{23}
+\frac{\mu}{\sqrt{2}}\,\,\frac{\pt u_2}{\pt a^D}\right|^2+
g^2_{{\rm unbr}}\left| 
\mu\,\frac{\pt u_2}{\pt a_{{\rm unbr}}}\right|^2
\nonumber\\[3mm]
&+&
2\left|\left(n_q \, a_q +\frac{m_A}{\sqrt{2}} \right) q^{1A}\right|^2 + 2\left|\left(n_q \, a_q +\frac{m_A}{\sqrt{2}} \right)
\bar{\tilde{q}}^{1A}\right|^2
\nonumber\\[3mm]
&+&
2\left|a^D M\right|^2 + 2\left|a^D\bar{\tilde{M}}\right|^2 +\cdots \, ,
\label{potr=1}
\eeqn
where $a_q$, $a^D$ and $a_{{\rm unbr}}$ are scalar superpartners of the gauge potentials  in (\ref{Aq}), (\ref{AM}) and
(\ref{Aunbr}), while $g_q$, $g_M$ and $g_{{\rm unbr}}$ are 
the corresponding U(1) gauge couplings. 
The dots represent the $D$ terms.
From Eq. (\ref{potr=1}) we learn that
\beqn
n_q \,\langle  \,\tilde{q}_{A1} q^{1A} \rangle & = &
-\frac{\mu}{\sqrt{2}}\,\,\frac{\pt u_2}{\pt a_q},
\nonumber\\[3mm]
\langle  \tilde{M}_{23} M_{23} \rangle & = &
-\frac{\mu}{\sqrt{2}}\,\,\frac{\pt u_2}{\pt a^D},
\nonumber\\[3mm]
\frac{\pt u_2}{\pt a_{{\rm unbr}}} & = & 0, 
\label{r=1vac}
\eeqn
while $a^D =0$ and $\sqrt{2}\,n_q\,a_q = - m_1$. All derivatives in  Eqs. (\ref{r=1vac}) can be calculated
from the Seiberg-Witten curve which factorizes in the $r=1$ vacuum at hand as follows:
\beq
y^2=(x-e_1)^2\,(x-e_2)^2\,(x-e_3^{+})(x-e_3^{-}).
\label{curveN=3}
\eeq
Double roots at $x=e_1$ and $x=e_2$ are associated with the light quark $q^{11}$ and light monopole $M_{23}$,
respectively. Details of this calculation can be found in Appendix. The result is
\beq
\langle  \tilde{q}_{11} q^{11} \rangle  = 
\frac{\xi_1}{2}, \qquad
\langle  \tilde{M}_{23} M_{23} \rangle  = 
\frac{\xi_2}{2}\,,
\label{qMvev}
\eeq
while the last equation in (\ref{r=1vac}) ensures that $e_3^{+}+e_3^{-}=0$, see (\ref{DijVafa}). Here $\xi_1$ and $\xi_2$ are given by (\ref{xiv}).

Again  we see that   all condensates, independently on their nature,  are determined by the same universal formula  
(\ref{xiv}). Above we analyzed only a few particular examples.  Extension to the general case is straightforward, however.

\section{Dynamical regimes and dualities in  \\ the \boldmath{$r$} vacua}
\label{five}

\subsection {Confinement and screening }
\label{conf}
\setcounter{equation}{0}

In the  hybrid $r$ vacua both quarks and monopoles charged with respect to orthogonal subgroups of U$(N)$ condense.
As a result, both the  non-Abelian magnetic strings \cite{HT1,ABEKY,SYmon,HT2}  and  the
Abelian Abrikosov-Nielsen-Olesen electric strings
develop supported by the quark and monopole condensates, respectively. 
Clearly, the magnetic strings confine monopoles while the
electric strings confine quarks. Now we focus on large quark masses,  with the quarks possessing  pure
color-electric  charges.\footnote{The
string formation and confinement in the $r$ dual theories at small quark masses due to  the 
quark-like dyon condensation
was studied in \cite{SYdual,SYrvacua}, see  Sec. \ref{sum}.}

Let us turn again to  the simplest example of the $r=1$ vacuum in the
U(3) gauge theory and show   how confinement and screening of different states  work in this case. A similar discussion for the $r=1$ vacuum in the SU(3) gauge theory can be found in \cite{MY}. 

All charges of condensed quark $q^{11}$ and monopole $M_{23}$ are given in Eq.~(\ref{charges}). Now we calculate the fluxes of the strings formed due to  condensation of these states. Consider first the magnetic strings. 

Since we have only one condensed quark $q^{11}$ in $r=1$ vacuum we deal with a single Abelian magnetic string, to be referred to as $S_m$. 
Suppose the $q^{11}$ quark  has a winding 
\beq
q^{11} \sim\sqrt{\frac{\xi_1}{2}}\,e^{i\alpha}, \qquad M_{23} \sim\sqrt{\frac{\xi_2}{2}}
\label{qwind}
\eeq
at $r\to\infty$
(see (\ref{qMvev})), where $r$ and $\alpha$ are the polar coordinates in the plane $i=1,2$ orthogonal 
to the string axis. 
Equations~(\ref{qwind})  imply the following behavior of the gauge potentials at $r\to\infty$:
\beqn
&& \frac12 A_i +\frac12 A_i^3 + \frac1{2\sqrt{3}} A_i^8  \sim \pt_i \alpha\,,
\nonumber\\[2mm]
&&  -\frac12 A_i^3 + \frac{\sqrt{3}}{2} A_i^8 \sim 0\,,
\label{qwindings}
\eeqn
as follows from the quark and monopole charges in (\ref{charges}). In the $r=1$ vacuum we have to supplement these conditions
with one extra condition  ensuring that the combination (\ref{Aunbr}) of 
the gauge potentials, which  interacts neither with the quark nor monopole, is not excited, namely,
\beq
 -\frac43 A_i + A_i^3 + \frac1{\sqrt{3}} A_i^8\sim 0\,.
\label{ort=0}
\eeq
The solution to  these equations is
\beqn
  A_i  \sim \frac67 \,\pt_i \alpha\,, \qquad
A_i^3  \sim \frac67\, \pt_i \alpha\,,\qquad
  A_i^8 \sim \frac{6}{7\sqrt{3}}\,\pt_i \alpha\,.
\label{gaugewindI}
\eeqn
It determines the  gauge  fluxes $\int dx_i A_i$, $\int dx_i A^3_i$ and  $\int dx_i A^8_i$ of the string $S_m$,
 respectively.
The integration above is performed over a large circle in the $(1,2)$ plane.

Next, we define the string charges \cite{SYdual} as 
\beqn
\int dx_i (A^D_i,\,A_i;\,A^{3D}_i,\,A^{3}_i;\,A^{8D}_i,\,A^{8}_i)
\equiv
 4\pi\,(-n_e,\,n_m;\,-n^3_e,\,n^3_m;-\,n^8_e,\,n^8_m)\,.\nonumber\\
\label{defstrch}
\eeqn
This definition guarantees that the string has the same charge as a probe monopole which can be attached to
the string endpoint. In other words, the flux of the given string is the flux of the probe
monopole sitting on string's end with
the charge defined by (\ref{defstrch}). Note, that this probe monopole does not necessarily exist in the
theory under consideration.  For example,  the  monopoles from the
SU$(r)$ sector are rather string junctions, so they are 
attached to two strings, \cite{SYmon,SYdual}. We will see below that the charges of
the physical monopoles confined in
the hybrid vacuum differ from the charge of the probe monopoles.

In particular, according to this definition, the charge of the string with the fluxes (\ref{gaugewindI})
is 
\beq
\vec{n}_{S_m}=\left(0,\,\frac37;\,0,\,\frac37;\,0,\,\frac3{7\sqrt{3}}\right).
\label{Sm}
\eeq
Since this string is associated with the quark winding, it is  magnetic. 

Now let us consider the electric string existing  due to the winding of the monopole $M_{23}$. In the vacuum at hand
we have
\beq
q^{11} \sim\sqrt{\frac{\xi_1}{2}}, \qquad M_{23} \sim\sqrt{\frac{\xi_2}{2}}\,e^{i\alpha}
\label{Mwind}
\eeq
at $r\to\infty$. Therefore, 
\beqn
   -\frac12 A_i^{3D} + \frac{\sqrt{3}}{2} A_i^{8D}\sim \pt_i \alpha\,,\qquad 
\frac12 A_i^{3D} + \frac1{2\sqrt{3}} A_i^{8D}  \sim 0\,.
\label{Mwindings}
\eeqn
Solution to these equation is
\beq
A_i^{3D}  \sim -\frac12 \,\pt_i \alpha\,, \qquad
A_i^{8D}  \sim \frac{\sqrt{3}}{2}\, \pt_i \alpha\,.
\eeq
The gauge potential $A_i^{D}$ is not excited.
This gives the charge of the $S_e$ string, 
\beq
\vec{n}_{S_e}=\left(0,\,0;\,\frac14,\,0;\,\,-\frac{\sqrt{3}}{4},\,0\right).
\label{Se}
\eeq
Since this string is associated with the monopole winding, it is  electric.  

\begin{figure}
\epsfxsize=8cm
\centerline{\epsfbox{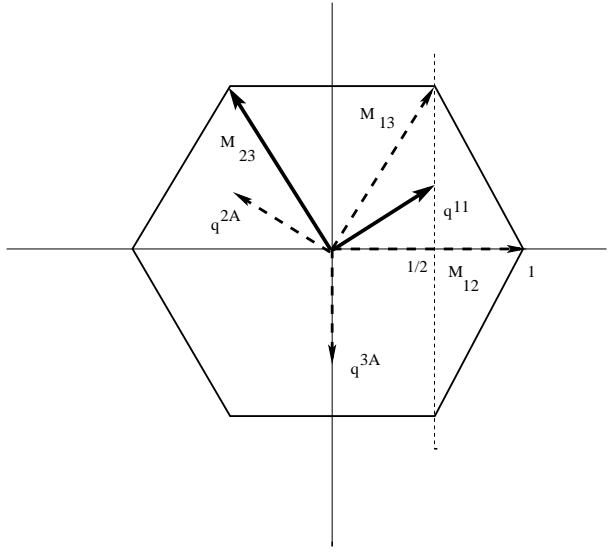}}
\caption{\small Projection of charges  of different quark and monopole states  to SU(3) subalgebra of U(3).
Charges of condensed states are shown by solid arrows, while charges of confined states are shown by dashed arrows.}
\label{fig:SU3charges}
\end{figure}

It is instructive to check that all quarks and elementary monopoles are either screened or confined in the
hybrid vacuum under consideration.
Clearly the quarks $q^{1A}$ and monopoles $M_{23}$ are screened. Let us analyze  other quarks
$q^{2A}$,  $q^{3A}$ as well as  monopoles $M_{12}$, $M_{13}$. The SU(3) projections of the charges of these states
are shown in Fig.~\ref{fig:SU3charges}. Note, that these states are heavy and are not included in the low-energy
theory.

Start with the quark $q^{2A}$. It should be confined by the electric sting $S_e$. It is not difficult to verify this.  Indeed,
the charge of this quark can be represented as
\beq
\vec{n}_{q^{2A}}=\left(\frac12,\,0;\,-\frac12, \,0;\,\frac1{2\sqrt{3}},0\,\right)=
-\vec{n}_{S_e} + \frac17\,\vec{n}_{q^{11}} + \frac97\,\vec{n}_{{\rm unbr}}^e,
\label{q2}
\eeq
where 
\beq
\vec{n}_{{\rm unbr}}^e=\left(\frac13,\,0;\,-\frac14, \,0;\,-\frac1{4\sqrt{3}},0\,\right)
\eeq
is the source for the electric U(1)$^{{\rm unbr}}$  gauge field (\ref{Aunbr}). This U(1) is unbroken.

We see that the $q^{2A}$ quark  is confined. Part of its electric flux is confined by the electric string (\ref{Se}).
Another part  is screened by the $q^{11}$ condensate. What is left is precisely the flux of  the unbroken gauge field
U(1)$^{{\rm unbr}}$. 

Of course, any three-dimensional vector of the quark $q^{2A}$ charges  can always be written as 
a linear combination of 
three orthogonal vectors. What is  nontrivial in Eq. (\ref{q2}), however, is the coefficient in front
of the string charge: it should be integer to ensure confinement.

\begin{figure}
\epsfxsize=8cm
\centerline{\epsfbox{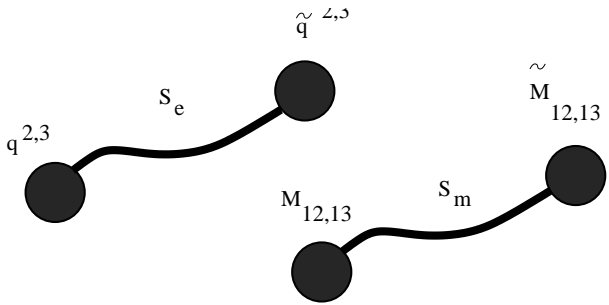}}
\caption{\small Stringy mesons made of quarks and monopoles.}
\label{fig:hybridconf}
\end{figure}

As a result of  confinement and screening stringy mesons made of  quarks and antiquarks $q^{2A}$ 
connected by  strings $S_e$ are formed, see Fig.~\ref{fig:hybridconf}. The string endpoints  
emit electric fluxes of the unbroken U(1)$^{{\rm unbr}}$. This makes this meson a dipole-like configuration, cf. \cite{SYrvacua}.
All other color fluxes are either confined or screened inside the meson.

Analogously we can convince ourselves that the quark $q^{3A}$ is confined too.
To check this we represent the charge of this quark as
\beq
\vec{n}_{q^{3A}}=\left(\frac12,\,0;\, 0,\,0;\,-\frac1{\sqrt{3}},0\,\right)=
\vec{n}_{S_e} + \frac17\,\vec{n}_{q^{11}} + \frac97\,\vec{n}_{{\rm unbr}}^e\,.
\label{q3}
\eeq
Thus, the $q^{3A}$ quark  is obviously confined by the electric string $S_e$. 
The unconfined part of its flux is 
screened by the $q^{11}$ condensate while the remainder coincides with the flux of unbroken U(1)$^{{\rm unbr}}$.

Now we will pass to confinement of the monopoles. Decomposing
\beq
\vec{n}_{M_{12}}=\left(0,\,0;\, 0,\,1;\,0,\,0\,\right)=
\vec{n}_{S_m} - \frac12\,\vec{n}_{M_{23}} - \frac97\,\vec{n}_{{\rm unbr}}^m
\label{M12}
\eeq
we see that the part of the monopole $M_{12}$ flux is confined by the magnetic string $S_m$ (see (\ref{Sm})), while the the second term  is screened by the $M_{23}$ condensate.
The remainder of the flux is proportional to 
\beq
\vec{n}_{{\rm unbr}}^m =\left(0,\,\frac13;\,0,\,-\frac14;\,0, \,-\frac1{4\sqrt{3}}\,\right),
\eeq
which  is the source for the  unbroken magnetic gauge field U(1)$^{{\rm unbr}}$.

As a result, a meson formed by the magnetic string $S_m$ with the $M_{12}$ monopole and its antimonopole
 attached to the endpoints appears in the physical spectrum. This meson is a dipole-like configuration emitting magnetic fluxes
of the unbroken gauge field U(1)$^{{\rm unbr}}$, see Fig.~\ref{fig:hybridconf}.

For the $M_{13}$ monopole  we have 
\beq
\vec{n}_{M_{13}}=\left(0,\,0;\, 0,\,\frac12;\,0,\,\frac{\sqrt{3}}{2}\,\right)=
\vec{n}_{S_m} + \frac12\,\vec{n}_{M_{23}} - \frac97\,\vec{n}_{{\rm unbr}}^m\,,
\label{M13}
\eeq
This monopole is apparently confined by the same  $S_m$ magnetic string.

Note, that in the simple case at hand ($r=1$) we have a single condensed quark  and a single condensed monopole
($N-r-1=1$). Therefore other (confined)  quarks and monopoles play a role of the endpoints of
electric and magnetic strings, respectively. In the  case of generic $r$, with $r$ condensed quarks, we have 
$r$ elementary magnetic non-Abelian strings. Hence,
the confined elementary monopoles of the SU($r$) subgroup become junctions of two ``neighboring'' strings
\cite{SYmon,SYrvacua}.
Similarly,  for a generic value of $(N-r-1)$ (i.e. $N-r-1$  condensed monopoles)  we have $(N-r-1)$ Abelian electric strings,
thus certain confined quarks become junctions of two different elementary electric strings \cite{HSZ}.

\subsection{\boldmath{$r$} Duality in \boldmath{\ntwo}}
\label{rduali}

In Sec. \ref{sum} we will briefly analyze various phases attainable in \ntwo SQCD in the limit of small quark masses.
It is instructive to discuss now the transition to this limit.

From Sec.~\ref{larger} we know that the low-energy theory in the $r$ vacuum with $r> N_f/2$ is at weak coupling because
the quark masses are large and hence $\sqrt{\xi}\gg \Lambda$. However, if we reduce 
the quark masses making the parameters $\xi$ small 
the quark  sector runs to strong coupling,
and the theory undergoes  a crossover transition.  

At small values of $\xi$
low-energy physics can be described by a dual weakly coupled infrared free  $r$-dual theory \cite{SYrvacua}. 
The gauge group of the $r$-dual theory is 
\beq
U(\nu)\times U(1)^{N-\nu}, \qquad
\nu=\left\{
\begin{array}{cc}
r, & r\le \frac{N_f}{2}\\[2mm]
N_f-r, & r > \frac{N_f}{2} \\
\end{array}
\right. .
\label{nu}
\eeq
The light matter sector
 of the $r$-dual theory is represented by $N_f$ flavors of non-Abelian quark-like dyons charged with respect to the gauge group  SU$(\nu)$ (as well as a combination of Abelian factors in 
(\ref{nu})), plus $(r-\nu)$ singlet quarks and $(N-r-1)$ monopoles charged with respect to different Abelian factors  in (\ref{nu}). The color charges of  the non-Abelian quark-like dyons are identical to those of quarks.\footnote{Because of monodromies, the quarks (preserving their 
 weight-like electric charges) pick up certain root-like magnetic charges at strong coupling.} However, they belong to a different representation of the global color-flavor locked group. VEVs of both non-Abelian quark-like
 dyons and quark singlets are still given by Eq. (\ref{xiv}) with $P=1,..., r$ \cite{SYrvacua}.
 
 Upon condensation of the quark-like dyons in the U$(\nu)$ sector of the $r$-dual theory
non-Abelian string are formed. These
strings still confine monopoles,  rather than quarks \cite{SYdual,SYrvacua}. Thus, $r$ duality is  not  electromagnetic. 

At strong coupling  where the dual description is applicable, the 
quarks and gauge bosons of the original theory from the U$(\nu)$ sector are in 
the ``instead-of-confinement'' phase. Namely,
the Higgs-screened quarks and
gauge bosons decay into monopole-antimonopole 
pairs on the curves of marginal stability (CMS) \cite{SYdual,SYtorkink}.
The (anti)monopoles  pair is confined. In other words, the original quarks and gauge bosons 
evolve at small  $\xi$  into monopole-antimonopole stringy mesons (presumably forming 
the Regge trajectories). 

Note, that the presence of the SU$(\nu)\times $U(1)$^{N_f-\nu}$ gauge groups at 
the roots of the Higgs branches in massless ($\xi=0$)  \ntwo SU$(N)$ SQCD was first 
recognized long ago in \cite{APS}, see also \cite{CKM}.

\subsection{Phases of \boldmath{\ntwo} SQCD at small masses}
\label{sum}

In this section we summarize for completeness the phases of $\mu$-deformed \ntwo QCD with small quark masses
(and small $\mu$). First, we will discuss the small -$r$ vacua, namely,   $r<N_f/2$. 

As we reduce the quark masses, the  quantum numbers of
the  light states change due to monodromies \cite{SW1,SW2,BF}. In particular,
the quarks pick up root-like color-magnetic charges in addition to their weight-like color-electric charges.
Still (in the  $r< N_f/2 $ vacua) there is no crossover, the low-energy theory remains the same:
infrared free U$(r)\times$U(1)$^{N-r}$ gauge theory with $N_f$ quarks (or, more exactly, what becomes of quarks)
and $N-r-1$ singlet monopoles \cite{MY2}. It is at weak coupling provided the parameters $\xi_P$ are small enough.

The quarks from the U$(r)$ sector and monopoles form the orthogonal U(1)$^{N-r}$ still develop VEVs 
determined by Eq. (\ref{xiv}). 
Physics of screening and confinement also remains intact at small $m_A$. Say, if a given monopole state 
(charged with respect to the Cartan generators of SU$(r)$) is confined by the quark condensation at large masses, this confinement property does not change when we follow this given state to the small mass domain,  although the quark color charges  change 
\cite{MY2}. If quarks are screened in the $r$ vacuum at large masses
 they (or what becomes of quarks) are still screened in the same vacuum in the limit of small masses.
 Monodromies are nothing other than the  relabeling of states, they do not change physics.

In the  $r$ vacua with $r>N_f/2$ physics is quite different, see \cite{SYdual,SYrvacua} and Sec. \ref{larger}  above.
With decreasing  $\xi$ the theory undergoes a crossover transition. At small $\xi$ physics can be described by
weakly coupled infrared free $r$-dual theory with the gauge group U$(\nu)\times$U(1)$^{N-\nu}$ and  $\nu=N_f-r$.
The quarks from U$(\nu)$ sector are in the ``instead-of-confinement'' phase: the Higgs-screened quarks decay into 
the monopole-antimonopole pairs confined by the non-Abelian strings. The singlet  quarks from the  U(1)$^{r-\nu}$ sector
and the monopoles from U(1)$^{N-r}$ sector are Higgs-screened. Other monopoles charged with respect to 
Cartan generators of SU$(r)$ and heavy quarks charged with respect to the orthogonal U(1)$^{N-r}$ are confined.

\section{Conclusions}

Our main result is the demonstration of the fact that 
VEVs of all condensates -- quark and monopole -- in the hybrid $r$ vacua of \ntwo SQCD are given by the 
unified exact formula (\ref{xiv}). In the limit of infinitely heavy quarks, when the theory under consideration becomes pure glue, this formula implies the will-known sine formula for the string tensions. (The $P$ strings appearing in 
(\ref{YMmvev}) are usually referred to as $k$ strings.)

In Sec. \ref{five} we briefly   discuss dynamical regimes and dualities in the hybrid  $r$ vacua. Due to the condensation of 
$r$ quarks and $(N-r-1)$ monopoles we have $r$ non-Abelian magnetic and $(N-r-1)$ Abelian Abrikosov-Nielsen-Olesen electric strings in such vacua.
Magnetic strings confine monopoles, while electric strings confine quarks. We calculate the fluxes of 
the confining strings. A similar discussion in the SU($N$) theory was presented in
\cite{MY}.

Dynamical regimes and their change crucially depend on the value of $r$.
In the $r< N_f/2$ vacua the small quark mass domain does not qualitatively differ from the large quark mass domain: confinement and screening are essentially the same. In $r> N_f/2$ vacua the physics is rather different.
With decreasing  $m_A$ (and hence decreasing $\xi$) the theory undergoes a crossover transition
and at small $\xi$ can be described using $r$ duality.

\section*{Acknowledgments}
This work  is supported in part by DOE grant DE-FG02-94ER40823. 
The work of A.Y. was  supported 
by  FTPI, University of Minnesota, 
by RFBR Grant No. 13-02-00042a 
and by Russian State Grant for 
Scientific Schools RSGSS-657512010.2.

\section*{Appendix:  The \boldmath{$r=1$} vacuum in U(3)}

 \renewcommand{\theequation}{A.\arabic{equation}}
\setcounter{equation}{0}

In this Appendix we calculate the derivatives $\pt u_2 /\pt a_q$ and $\pt u_2 /\pt a_D$ which appear
in the right-hand sides of 
Eqs. (\ref{r=1vac}) for the quark and monopole condensates in the $r=1$ vacuum of the U(3) theory.
This calculation is quite similar to the calculation in the  $r=0$ vacuum in the U(2) theory in Sect. \ref{monopolevac} 
and in the $r=3$ vacuum in the U(3) theory in \cite{SYfstr}. Therefore, we will be brief.

\begin{figure}
\epsfxsize=7cm
\centerline{\epsfbox{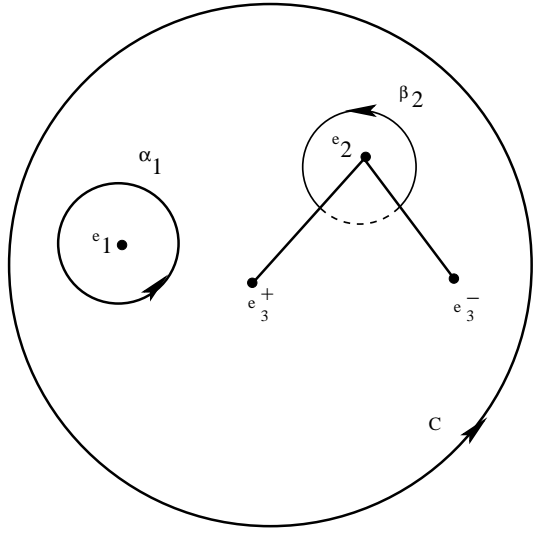}}
\caption{\small $\alpha_1$, $\beta_2$ and $C$-contours in $x$-plane for the U(3) theory. Solid straight lines denote  cuts. }
\label{fig:r=1contours}
\end{figure}

Explicit expressions from \cite{ArFa,KLTY,ArPlSh,HaOz}  generalized to the 
 U$(N)$ case \cite{SYfstr} imply
\beqn
&&
 \frac{\pt \Phi_1}{\pt u_k}  =  \,\frac1{2\pi i}\oint_{\alpha_1}  \frac{dx}{y}\,P_k (x) + \delta_{k1},
\nonumber\\[3mm]
&&
\frac{\pt a^D}{\pt u_k}  =  \,\frac1{2\pi i}\oint_{\beta_2}  \frac{dx}{y}\,P_k (x) + \delta_{k1},
\nonumber\\[3mm]
&&
\frac{\pt \left(\Phi_1+\Phi_2+\Phi_3\right)}{\pt u_k}  =  \,\frac1{2\pi i}\oint_{C}  \frac{dx}{y}\,P_k (x) + 3\delta_{k1},
\label{dadu31}
\eeqn
where $\Phi_1$, $\Phi_2$ and $\Phi_3$ are diagonal elements of the matrix 
$\Phi$, see (\ref{Phi}), while the polynomials $P_k (x)$, $k=1,2,3$
are given by
\beqn
P_3 (x)
&= & \frac13
 \nonumber\\[3mm]
P_2 (x)
&=&
 \frac12 \,\left[x-\frac13\left(e_1+e_2+e_3\right)\right],
\nonumber\\[3mm]
P_1 (x)
&=&
 - 2\left[x^2-\frac12 x\left(e_1+e_2+e_3\right)+
\frac19 \left(e_1+e_2+e_3\right)^2\right]
\label{dadu32}
\eeqn
and $e_3 =(e_3^{+} + e_3^{-})/2$.
Here the contours $\alpha_1$ and $\beta_2$ encircle 
the double roots $e_1$ and $e_2$ of the Seiberg-Witten curve (\ref{curveN=3}) 
associated with the  light quark $q^{11}$ and the light monopole $M_{23}$, respectively, while $C$ is the contour at infinity, see Fig.~\ref{fig:r=1contours}. 

The contour integrals in  (\ref{dadu31}) can be readily calculated, in particular the integrals along the contours $\alpha_1$ and $\beta_2$ are given by their pole contributions. These integrals  determine 
the derivatives of $a_q$ and $a_{{\rm unbr}}$ with respect to
$u_k$ since $\Phi_1=n_q\, a_q$, while  $a_{{\rm unbr}}$
is a linear combination of  of $a_q$ and $\left(\Phi_1+\Phi_2+\Phi_3\right)= 3a/2$, see Eq. (\ref{Aunbr}).
Inverting the matrix $\pt (a_q, a^D,a_{{\rm unbr}} ) / \pt u_k $ we get the desired expressions
for $ \pt u_2 /\pt a_q $, $ \pt u_2 /\pt  a^D $
and $ \pt u_2 /\pt a_{{\rm unbr}} $ in terms of the roots of the Seiberg-Witten curve.

Ommiting details presented in Sect. \ref{monopolevac} 
and  \cite{SYfstr} for similar cases we arrive at the results for the quark and monopole VEVs quoted in Eq. (\ref{qMvev}). Also, the last equation in (\ref{r=1vac}) gives $e_3=0$, in accordance with (\ref{DijVafa}).

\end{document}